\documentclass[preprintnumbers,showpacs,amsmath,amssymb,floatfix,superscriptaddress,nofootinbib]{revtex4}
\usepackage{graphicx}
\usepackage{epsfig}
\usepackage{bm}
\usepackage{amsfonts}

\def\rl{\rho_\Lambda}
\def\ol{\Omega_\Lambda}

\begin{document}

\title{Observational constraints on holographic dark energy with varying gravitational constant}

\author{Jianbo Lu}
\email{lvjianbo819@163.com}
\affiliation{Institute of Theoretical Physics, School of Physics \&
Optoelectronic Technology, Dalian University of Technology, Dalian,
116024, P. R. China}

\author{Emmanuel N. Saridakis}
 \email{msaridak@phys.uoa.gr}
 \affiliation{College of Mathematics
and Physics,\\ Chongqing University of Posts and
Telecommunications, Chongqing, 400065, P.R. China }

\author{M. R. Setare}
\email{rezakord@ipm.ir} \affiliation{Department of Science, Payame
Noor University, Bijar, Iran }

\author{Lixin Xu}
\email{lxxu@dlut.edu.cn} \affiliation{Institute of Theoretical
Physics, School of Physics \& Optoelectronic Technology, Dalian
University of Technology, Dalian, 116024, P. R. China}

\begin{abstract}
We use observational data from  Type Ia Supernovae (SN), Baryon
Acoustic Oscillations (BAO), Cosmic Microwave Background (CMB) and
observational Hubble data (OHD), and the Markov Chain Monte Carlo
(MCMC) method, to constrain the cosmological scenario of
holographic dark energy with varying gravitational constant. We
consider both flat and non-flat background geometry, and we
present the corresponding constraints and contour-plots of the
model parameters. We conclude that the scenario is compatible with
observations. In 1$\sigma$ we find
$\Omega_{\Lambda0}=0.72^{+0.03}_{-0.03}$,
$\Omega_{k0}=-0.0013^{+0.0130}_{-0.0040}$,
$c=0.80^{+0.19}_{-0.14}$ and $\Delta_G\equiv
G'/G=-0.0025^{+0.0080}_{-0.0050}$,  while for the present value of
the dark energy equation-of-state parameter we obtain
$w_0=-1.04^{+0.15}_{-0.20}$.
\end{abstract}

\pacs{95.36.+x, 98.80.-k, 98.80.Es}
 \maketitle

\section{Introduction}

Nowadays it is strongly believed that the universe is experiencing
an accelerated expansion, and this is supported by many
cosmological observations, such as SNe Ia \cite{1}, WMAP \cite{2},
SDSS \cite{3} and X-ray \cite{4}. A first direction that could
provide an explanation of this remarkable phenomenon is to
introduce the concept of dark energy, with the most obvious
theoretical candidate being the cosmological constant. However, at
least in an effective level, the dynamical nature of dark energy
can also originate from   a variable cosmological ``constant''
\cite{varcc}, or form various fields, such is a canonical scalar
field (quintessence) \cite{quint}, a phantom field, that is a
scalar field with a negative sign of the kinetic term
\cite{phant}, or the combination of quintessence and phantom in a
unified model named quintom \cite{quintom}. The second direction
that could explain the acceleration is to modify the gravitational
theory itself, such is the generalization to $f(R)$-gravity
\cite{12}), scalar-tensor theories with non-minimal coupling
\cite{9}, string-inspired models \cite{11} etc.

Going beyond the aforementioned effective description requires a
deeper understanding of the underlying theory of quantum gravity
unknown at present. However, physicists can still make some
attempts to probe the nature of dark energy according to some
basic quantum gravitational principles. Currently, an interesting
such an attempt is the so-called ``Holographic Dark Energy''
proposal \cite{Hsu:2004ri,Li:2004rb}. Its framework is the black
hole thermodynamics  and the connection (known from AdS/CFT
correspondence) of the UV cut-of of a quantum field theory, which
gives rise to the vacuum energy, with the largest distance of the
theory \cite{Cohen:1998zx}. Thus, determining an appropriate
quantity $L$ to serve as an IR cut-off, imposing the constraint
that the total vacuum energy in the corresponding maximum volume
must not be greater than the mass of a black hole of the same
size, and saturating the inequality, one identifies the acquired
vacuum energy as holographic dark energy:
\begin{equation}\label{de}
\rho_\Lambda=\frac{3c^2}{8\pi G L^2},
\end{equation}
with $G$ the Newton's gravitational constant and $c$ a constant.
The holographic dark energy scenario has been tested and
constrained by various astronomical observations \cite{obs3a} and
it has been extended to various frameworks
\cite{nonflat,holoext,intde}.

However, there are  indications that Newton's ``constant'' $G$ can
by varying, being a function of time or equivalently of the scale
factor \cite{G4com}. In particular, observations of Hulse-Taylor
binary pulsar \cite{Damour,kogan}, helio-seismological data
\cite{guenther}, Type Ia supernova observations \cite{1}  and
astereoseismological data from the pulsating white dwarf star
G117-B15A \cite{Biesiada} lead to $\left|\dot{G}/G\right|
\lessapprox 4.10 \times 10^{-11} yr^{-1}$, for $z\lesssim3.5$
\cite{ray1}. Thus, in our previous paper \cite{Jamil:2009sq}, we
investigated the holographic dark energy scenario under a varying
gravitational constant, and we extracted the corresponding
corrections to the dark energy equation-of-state parameter.

In the present work we are interested in constraining the model of
holographic dark energy with a varying Newton's constant, using
various observational data arising from  SN, BAO, CMB and OHD.
Such an analysis is crucial for the validity of the aforementioned
scenario. The plan of the work is as follows: In section
\ref{model} we present briefly the holographic dark energy
scenario with a varying Newton's constant and the corresponding
expressions of the dark-energy equation-of-state parameter. In
section \ref{obsconstr} we perform a combined observational
constraint analysis, allowing for variations   in all
  model parameters. Finally, section
\ref{Conclusions} is devoted to the summary of our results.

\section{Holographic dark energy with varying gravitational constant}
\label{model}

In this section we briefly review the holographic dark energy
proposal, in the case where the gravitational constant is itself a
function of the scale factor \cite{Jamil:2009sq}. In order for our
results to be more transparent we examine separately the flat and
non-flat background geometries.

\subsection{Flat universe}

In the case where the space-time geometry is a flat
Robertson-Walker:
\begin{equation}\label{met}
ds^{2}=-dt^{2}+a(t)^{2}(dr^{2}+r^{2}d\Omega^{2}),
\end{equation}
with $a(t)$ the scale factor and $t$ the comoving time. As usual,
the first Friedmann equation reads:
\begin{equation}\label{FR1}
H^2=\frac{8\pi G}{3}\Big(\rho_m+\rl\Big),
\end{equation}
with $H$ the Hubble parameter, $\rho_m=\frac{\rho_{m0}}{a^3}$,
where $\rho_m$ and $\rl$ stand respectively for  matter and dark
energy densities and the index $0$ marks the present value of a
quantity. Furthermore, we will use the density parameter $
\ol\equiv\frac{8\pi G}{3H^2}\rl$, which, imposing explicitly the
holographic nature of dark energy according to relation
(\ref{de}), becomes
\begin{eqnarray}
 \label{OmegaL2}
\ol=\frac{c^2}{H^2L^2}.
\end{eqnarray}
As usual, in the case of a flat universe, the best choice for the
definition of $L$ is to identify it with the future event horizon
\cite{Li:2004rb,Guberina,Hsu:2004ri}, that is $L\equiv R_ h(a)$
with
\begin{equation}
 R_ h(a)=a\int_t^\infty{dt'\over
a(t')}=a\int_a^\infty{da'\over Ha'^2}~.\label{eh}
\end{equation}
 In this case, using  (\ref{OmegaL2}) and the
Friedmann equation (\ref{FR1}), one can show that
\begin{equation}\label{OmegaLdif3}
\ol'=\ol(1-\ol)\Big[1+\frac{2\sqrt{\ol}}{c}\Big]-\ol(1-\ol)\Delta_G,
\end{equation}
where primes denote the derivatives with respect to
 $\ln a$. In this expression the first term is the usual holographic dark energy differential
equation \cite{Li:2004rb}, while the second term is the correction
arising from the varying nature of $G$, which is quantified by the
parameter $\Delta_G\equiv G'/G$. Finally, concerning the
dark-energy equation-of-state parameter $w$, considered as a
function of the redshift $z$, we use the approximation
\cite{Chevallier:2000qy}
\begin{equation}
\label{w01flat}
 w(z)\approx w_0+w_1\left(\frac{z}{1+z}\right),
\end{equation}
with \cite{Jamil:2009sq}
\begin{eqnarray}
&&w_0=-{1\over 3}-{2\over 3c}\sqrt{\Omega_{\Lambda0}}
+\frac{\Delta_G}{3}\label{w0fl}\\
\label{w1fl}
 &&w_1={1\over
6c}\sqrt{\Omega_{\Lambda0}}(1-\Omega_{\Lambda0})\left(1+{2\over
c}\sqrt{\Omega_{\Lambda0}}\right)
-\frac{(1-\Omega_{\Lambda0})\sqrt{\Omega_{\Lambda0}}}{6c}\Delta_G.
\ \ \ \ \ \ \   \ \ \ \ \
\end{eqnarray}
In these expressions, the index $0$ stands for the value of a
quantity at present, where the Hubble parameter is $H_0$ and the
scale factor is $a_0=1$.

In summary, using the results of this subsection, and the usual
relations $\Omega_m=\Omega_{m0}a^{-3}$ (with
$\Omega_m\equiv\frac{8\pi G}{3H^2}\rho_m$ the matter density
parameter) and $\Omega_\Lambda=\Omega_{\Lambda0}a^{-3(1+w)}$,  we
can write the Friedmann equation (\ref{FR1}) in a form suitable
for observational elaboration as:
\begin{equation}\label{FR1b}
H^2=H_0^2\left\{\Omega_{m0}(1+z)^3+\Omega_{\Lambda0}(1+z)^{3[1+
w(z)]}\right\}.
\end{equation}

\subsection{Non-flat universe}

Let us now generalize the above result in the case of a general
FRW universe with line element
\begin{equation}\label{metr}
 ds^{2}=-dt^{2}+a^{2}(t)\left(\frac{dr^2}{1-kr^2}+r^2d\Omega^{2}\right)
\end{equation}
in comoving coordinates  $(t,r,\theta,\varphi)$, where $k$ denotes
the spacial curvature with $k=-1,0,1$ corresponding to open, flat
and closed universe respectively. The first Friedmann equation
writes:
\begin{equation}\label{FR1nf}
H^2+\frac{k}{a^2}=\frac{8\pi G}{3}\Big(\rho_m+\rl\Big).
\end{equation}
In this case, the cosmological length $L$ in (\ref{OmegaL2}) is
considered to be  \cite{nonflat}:
\begin{equation}\label{Lnonflat}
L\equiv\frac{a(t)}{\sqrt{|k|}}\,\text{sinn}\left(\frac{\sqrt{|k|}R_h}{a(t)}\right),
\end{equation}
where
\begin{equation}\frac{1}{\sqrt{|k|}}\text{sinn}(\sqrt{|k|}y)=
\begin{cases} \sin y  & \, \,k=+1,\\
             y & \, \,  k=0,\\
             \sinh y & \, \,k=-1.\\
\end{cases}
\end{equation}
A straightforward calculation using (\ref{OmegaL2}) and
(\ref{FR1nf})
 leads to \cite{Jamil:2009sq}:
\begin{eqnarray}
\label{Omprimenf}
\Omega_\Lambda^\prime=\Omega_\Lambda\!\left[\!1\!-\!\Omega_k\!-\!\Omega_\Lambda\!+\!\frac{2\sqrt{\Omega_\Lambda}}{c}\,
\text{cosn}\!\left(\frac{\sqrt{|k|}R_h}{a}\right)\!(1-\Omega_\Lambda)\right]
-\Omega_\Lambda(1-\Omega_k-\ol)\frac{G^\prime}{G}.
\end{eqnarray}
where
\begin{equation}\text{cosn}(\sqrt{|k|}y)=
\begin{cases} \cos y  & \, \,k=+1,\\
             1 & \, \,  k=0,\\
             \cosh y & \, \,k=-1.\\
\end{cases}\end{equation}
In expression (\ref{Omprimenf}) we have also introduced the
curvature density parameter $\Omega_k\equiv-\frac{k}{(aH)^2}$. We
mention that since in this work we focus on observational
constraints, we have adopted the minus-sign convention for
$\Omega_k$-definition, which is the usual one in observational
works. Thus, in (\ref{Omprimenf}) $\Omega_k$ has the opposite sign
comparing to
 \cite{Jamil:2009sq}.
 Clearly, for $k=0$ (and thus
$\Omega_k=0$) it results to (\ref{OmegaLdif3}).
 Finally, concerning the
dark-energy equation-of-state parameter $w(z)$ we acquire $
w(z)\approx w_0+w_1\left(\frac{z}{1+z}\right) $, with
\cite{Jamil:2009sq}{\small{
\begin{eqnarray}\label{w0nonflb}
&&w_0=-{1\over 3}-{2\over
3c}\sqrt{\Omega_{\Lambda0}+c^2\Omega_{k0}}
+\frac{\Delta_G}{3}\\
\label{w1nonflb}
&&w_1=-\frac{\Omega_{k0}}{3}+\frac{1}{6c}\sqrt{\Omega_{\Lambda0}+c^2\Omega_{k0}}
\Big[1-\Omega_{k0}-\Omega_{\Lambda0}+\frac{2}{c}\left(1-\Omega_{\Lambda0}\right)\sqrt{\Omega_{\Lambda0}+c^2\Omega_{k0}}\Big]-
\frac{1}{6c}\sqrt{\Omega_{\Lambda0}+c^2\Omega_{k0}}
\left(1-\Omega_{k0}-\Omega_{\Lambda0}\right)\Delta_G.\ \ \ \ \ \
\end{eqnarray}
}}

In summary, using the results of this subsection,  we can write
the Friedmann equation (\ref{FR1nf}) in a form suitable for
observational elaboration as:
\begin{eqnarray}\label{FR1nfb}
H^2=H_0^2\left\{\Omega_{k0}(1+z)^2+\Omega_{m0}(1+z)^3+\Omega_{\Lambda0}(1+z)^{3[1+
w(z)]}\right\}.
\end{eqnarray}

\section{Observational constraints}
\label{obsconstr}

Let us now proceed to  a combined observational constraint
analysis of the scenario at hand, allowing for variations  in all
the aforementioned model parameters. We use observational data
from Type Ia Supernovae (SN), Baryon Acoustic Oscillations (BAO),
Cosmic Microwave Background $[l_A(z_\ast), R(z_\ast), z_{\ast}]$
(CMB) and Observational Hubble Data (OHD). The precise methods are
summarized in the Appendix.

In our calculations we take the total likelihood $L\propto
e^{-\chi^2/2}$ to be the product of the separate likelihoods of
SN, BAO, CMB and OHD. Thus, the $\chi^2$ is
\begin{eqnarray}
\chi^2(p_s)=\chi^2_{SN}+\chi^2_{BAO}+\chi^2_{CMB}+\chi^2_{OHD},
\end{eqnarray}
and the parameter vector reads
\begin{equation}
p_s=\{\Omega_{b}h^2, \Omega_{c}h^2, \Omega_k, c, \Delta_G\}.
\end{equation}
In addition, we obtain the three derived parameters
$\Omega_{\Lambda 0}$, $\Omega_{m0}=\Omega_{b}+\Omega_{c}$ and the
Hubble constant $H_0$, based on the above basic cosmological
parameters. In our analysis, we perform a global fitting on
determining the cosmological parameters using the Markov Chain
Monte Carlo (MCMC) method. The MCMC method is based on the
publicly available {\bf CosmoMC} package \cite{ref:MCMC}, which
has been modified to include the codes about BAO, CMB
$[l_A(z_\ast), R(z_\ast), z_{\ast}]$ and OHD. Finally, apart from
these and the two independent model parameters $c$ and $\Delta_G$,
the basic cosmological parameters are taken in the following
priors: the present physical baryon density
$\Omega_{b}h^2\in[0.005,0.9]$, the present physical cold dark
matter energy density $\Omega_{c}h^2\in[0.01,0.99]$, and for the
non-flat case, the additional parameter $\Omega_k\in[-0.1,0.1]$.

Using these techniques  we are able to impose constraints on the
various parameters of the scenario of holographic dark energy with
varying gravitational constant.

\subsection{Flat universe}

\begin{widetext}
\begin{figure}[ht]
\begin{center}
\includegraphics[width=19cm]{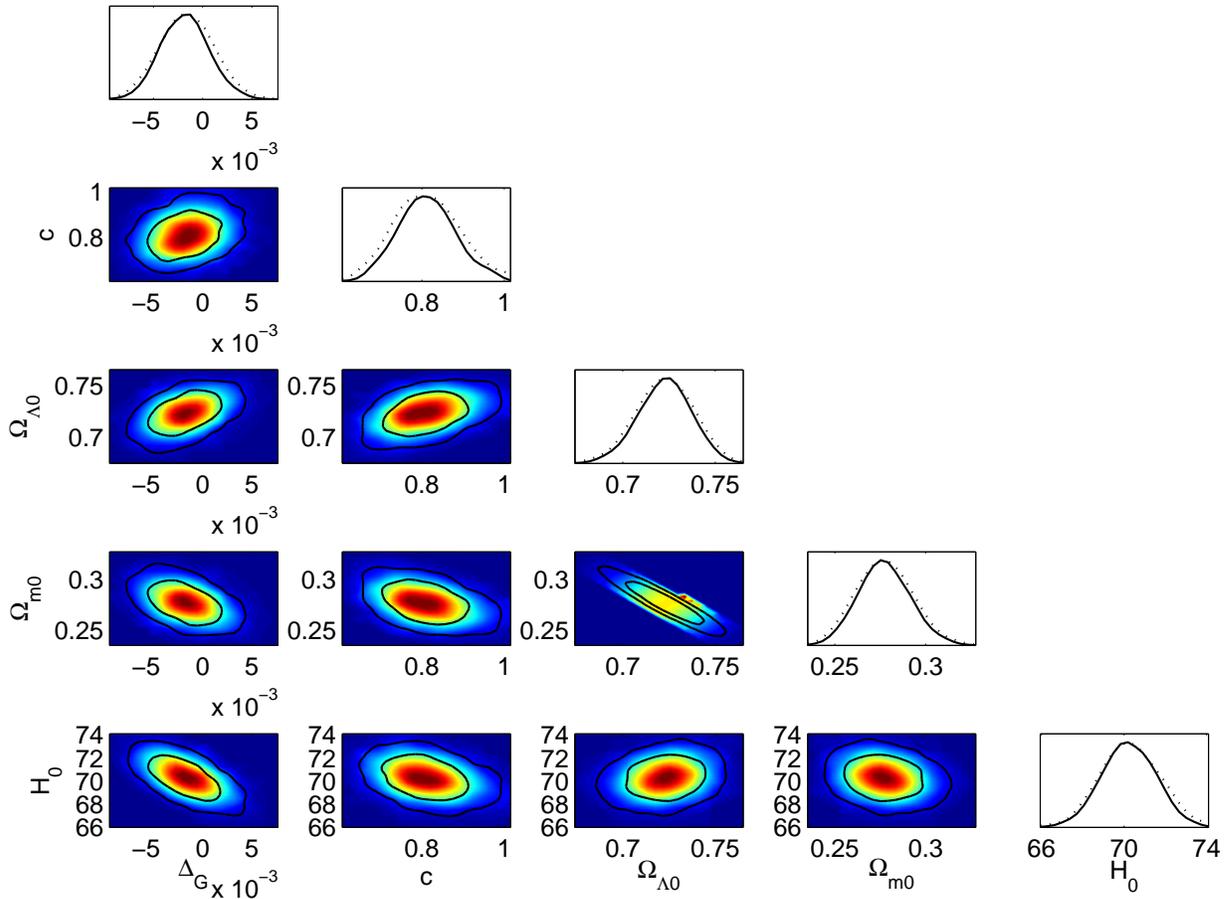}
\caption{ (Color online){\it{ The $1D$ and $2D$ likelihood plots,
    in the flat-background case, using SN, BAO, CMB and OHD observational data. The curves
    stand for the $1\sigma$ and $2\sigma$ regions. All quantities are dimensionless,
    apart from $H_0$ which is measured in $\rm km~s^{-1}\,Mpc^{-1}$. }}} \label{fig:flat}
\end{center}
\end{figure}
\end{widetext}

For the case of flat background geometry the cosmologically
interesting parameters are the dimensionless quantities
$\Omega_{\Lambda0}$, $c$ and $\Delta_G\equiv G'/G$, while we have
to take into account the additional uncertainty in $H_0$. The
corresponding $1D$ and $2D$ likelihood-contours are depicted in
Fig. \ref{fig:flat}. We mention that for completeness we provide
these plots not only for the parameters used in the fits, as
discussed above, but also for the derived ones, although some of
them are related to each other.
Additionally, in order to provide the results in a more
transparent way, in Table \ref{tab:flatcase} we present the
$1\sigma$
 best-fit values of the, used in the fits and derived ones,
model parameters.
\begin{table}[ht]
\begin{center}
\begin{tabular}{c|c|c|c|c}
\hline\hline
 $\chi^2_{min}$ & $\Omega_{\Lambda0}(1\sigma)$ & $c (1\sigma)$ &
 $\Delta_G(1\sigma)$ & $H_0(1\sigma)$\\ \hline
 $477.5$  & $0.723^{+0.026}_{-0.030}$ & $0.80^{+0.16}_{-0.13}$ & $-0.0016^{+0.0049}_{-0.0049}$ & $70.2^{+2.7}_{-2.5}$\\
\hline
\end{tabular}
\caption{The minimum value of $\chi^2$ and the 1$\sigma$  best-fit
values of the model parameters, in the flat-background case. All
quantities are dimensionless, apart from $H_0$ which is measured
in $\rm km~s^{-1}\,Mpc^{-1}$.}\label{tab:flatcase}
\end{center}
\end{table}

These figures show that the scenario at hand can be compatible
with observations. Furthermore, note that $\Delta_G\equiv G'/G$ is
restricted around zero, in a region which is in agreement with
independent observations and estimation of $G'/G$. In particular,
observations of Hulse-Taylor binary pulsar B$1913+16$ lead to the
estimation $\dot{G}/G\sim2\pm4\times10^{-12}{yr}^{-1}$
\cite{Damour,kogan}, while helio-seismological data provide the
bound $-1.6\times10^{-12}{yr}^{-1}<\dot{G}/G<0$ \cite{guenther}.
Similarly,  Type Ia supernova observations   give the best upper
bound of the variation of $G$ as $-10^{-11} yr^{-1} \leq
\frac{\dot G}{G}<0$   \cite{Gaztanaga}, while astereoseismological
data from the pulsating white dwarf star G117-B15A lead to
$\left|\frac{\dot G}{G}\right| \leq 4.10 \times 10^{-11} yr^{-1}$
\cite{Biesiada}. (See also \cite{ray1} for various bounds on
$\dot{G}/G$.) Since the limits of $G$-variation are given for
$\dot{G}/G$ in units $yr^{-1}$, and since $\dot{G}/G=H G'/G$, we
can estimate their implied $\Delta_G $ substituting the value of
$H$ in $yr^{-1}$. Thus, inserting an average estimation for the
Hubble parameter $H\approx\langle H\rangle\approx 6 \times
10^{-11} yr^{-1}$ \cite{Zhang:2009ae}, we obtain that
$0<|\Delta_G|\lesssim0.08$ \cite{Jamil:2009sq}. In summary, as we
observe from Fig. \ref{fig:flat}, $\Delta_G$ at 2$\sigma$ is well
inside these bounds, and this offers a self-consistency test for
our analysis.

For completeness, we close this subsection by estimating the
$1\sigma$ bounds of the dark energy equation-of-state parameter.
In particular, since we have approximated
 it as $w(z)\approx w_0+w_1\left(\frac{z}{1+z}\right)$, with $w_0$ and $w_1$ given by
 (\ref{w0fl}),(\ref{w1fl}), we can easily calculate that in
 1$\sigma$:
\begin{eqnarray}
&&w_0=-1.04^{+0.13}_{-0.12}\nonumber\\
&&w_1=0.15^{+0.04}_{-0.03}.
\end{eqnarray}
Interestingly enough, we observe that in this scenario the
best-fit value of the present value of $w$, namely $w_0$, is
smaller than the corresponding one of simple holographic dark
energy model \cite{Li:2004rb}, and it lies in the phantom regime.
Furthermore, note that the left and right bounds are larger as
expected, since we have the extra freedom in varying the
gravitational constant. Finally, $w_1$ is slightly larger
comparing to simple holographic dark energy scenario.

\subsection{Non-flat universe}

\begin{widetext}
\begin{figure}[ht]
\begin{center}
\includegraphics[width=19.3cm]{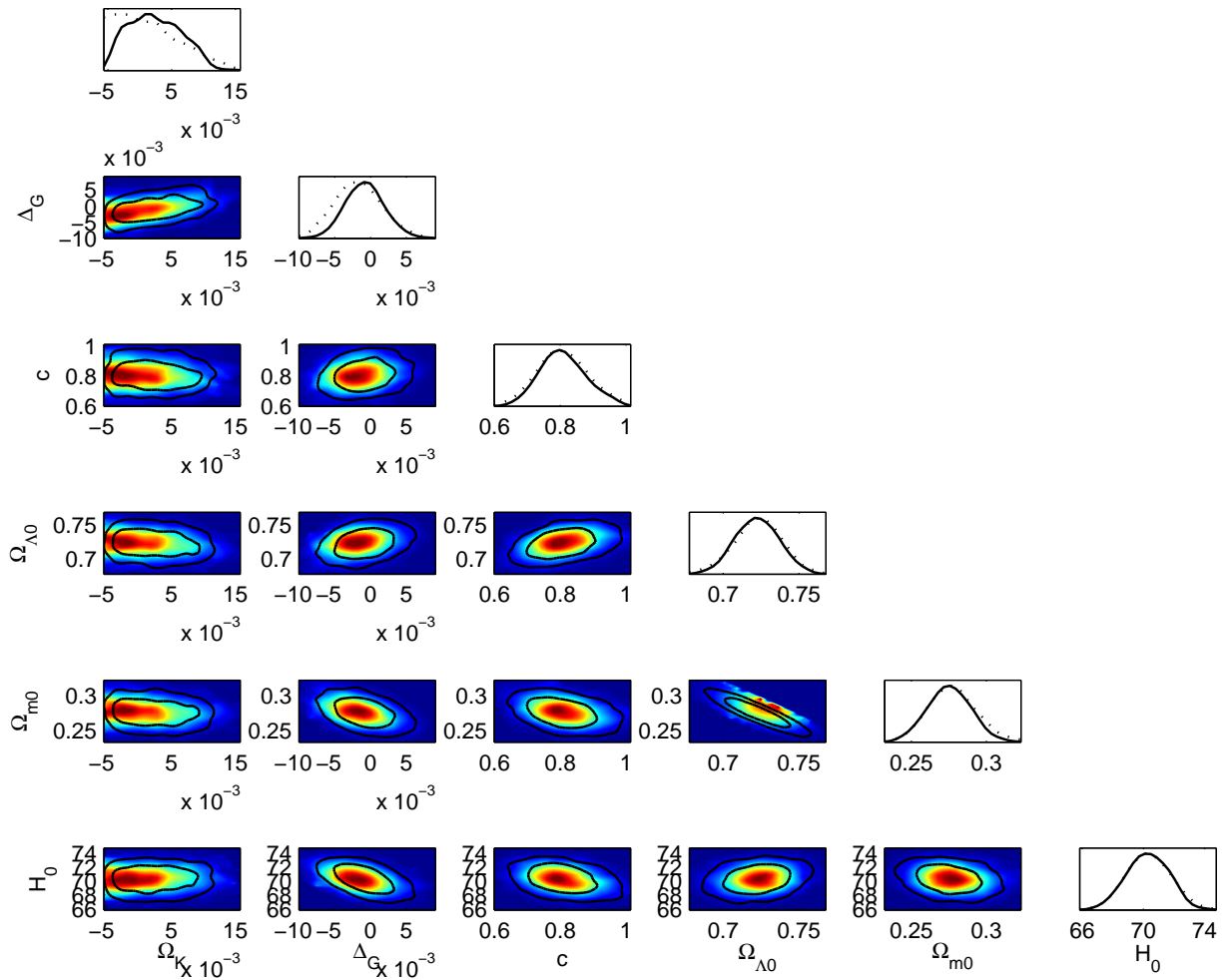}
\caption{ (Color online){\it{ The $1D$ and $2D$ likelihood plots,
    in the non-flat scenario, using SN, BAO, CMB and OHD observational data. The curves
    stand for the $1\sigma$ and $2\sigma$ regions. All quantities are dimensionless, apart from $H_0$ which is measured in $\rm km~s^{-1}\,Mpc^{-1}$.
 }}} \label{fig:nonflat}
\end{center}
\end{figure}
\end{widetext}

We use the combination of observational data from SNIa, BAO and
CMB to construct the likelihood contours for the free model
parameters, which in this case are the dimensionless quantities
$\Omega_{\Lambda0}$, $c$, $\Delta_G$ and $\Omega_{k0}$, together
with $H_0$. In Fig.~\ref{fig:nonflat} we present the corresponding
$1D$ and $2D$ likelihood-contours, not only for the parameters
used in the fits,  but also for the derived ones. Additionally, in
Table \ref{tab:non_flatcase} we present the 1$\sigma$  best-fit
values of the model parameters.
\begin{widetext}
\begin{table}[tbh]
\begin{center}
\begin{tabular}{c|c|c|c|c|c}
\hline\hline
 $\chi^2_{min}$ & $\Omega_{\Lambda0}(1\sigma)$ & $c (1\sigma)$ &
 $\Delta_G(1\sigma)$ & $\Omega_{k0}(1\sigma)$ & $H_{0}(1\sigma)$\\ \hline
 $477.4$  & $0.72^{+0.03}_{-0.03}$ & $0.80^{+0.19}_{-0.14}$ & $-0.0025^{+0.0080}_{-0.0050}$ & $-0.0013^{+0.0130}_{-0.0040}$ & $70.4^{+3.0}_{-2.9}$\\
\hline
\end{tabular}
\caption{The minimum value of $\chi^2$ and the 1$\sigma$ best-fit
values of the model parameters, in the non-flat scenario. All
quantities are dimensionless, apart from $H_0$ which is measured
in $\rm km~s^{-1}\,Mpc^{-1}$.}\label{tab:non_flatcase}
\end{center}
\end{table}
\end{widetext}

As we observe, the scenario of holographic dark energy with
varying gravitational constant in a non-flat background geometry,
is compatible with observations. As expected, the extra freedom in
varying the gravitational constant leads to larger bounds of all
the parameters, comparing to those of usual holographic dark
energy models \cite{obs3a}. However, as we see from the above
analysis, the variation range for $\Omega_{k0}$ is relatively
narrow comparing to the physically interesting range arising from
the usual investigations of the literature \cite{ref:Komatsu2008}.
This feature has to be considered as a disadvantage of the present
scenario.

Since we have approximated the dark energy equation-of-state
parameter as $w(z)\approx w_0+w_1\left(\frac{z}{1+z}\right)$, with
$w_0$ and $w_1$ given by
 (\ref{w0nonflb}),(\ref{w1nonflb}), we can easily estimate that in
 1$\sigma$:
\begin{eqnarray}
&&w_0=-1.04^{+0.15}_{-0.20}\nonumber\\
&&w_1=0.15^{+0.05}_{-0.10}.
\end{eqnarray}
As we see,  the present value of $w$, namely $w_0$, can lie in the
phantom regime. Additionally, note that the best-fit values are
very close to those of the flat scenario, since the best-fit value
for $\Omega_{k0}$ is very close to zero. However, the left and
right bounds are larger as expected. Finally, note that the
best-fit  values of both $w_0$ and $w_1$ are in very good
agreement with the corresponding ones of usual (constant-$G$)
holographic dark energy models \cite{obs3a}, but the bounds are
larger due to the extra freedom in the $G$-variation.

\section{Conclusions}
\label{Conclusions}

In this work we confronted the scenario of holographic dark energy
with varying gravitational constant, with data from SNIa, CMB, BAO
and OHD observations. We performed independent fittings for flat
and non-flat background geometries, and we extracted the
corresponding observational constraints on the free model
parameters of each case.

In the case of a flat universe, we deduced that the scenario at
hand is compatible with observations,  and we presented the
corresponding contour-plots on the free parameters, namely
  $\Omega_{\Lambda0}$, $c$ and
$\Delta_G$, taking into account the uncertainty in $H_0$. The
gravitational constant variation is inside the corresponding
bounds acquired from independent estimations of the literature,
and this acts as a self-consistency test of the model. The
corresponding 1$\sigma$-bounds are presented in Table
\ref{tab:flatcase}. Moreover, the best-fit  value of the present
dark energy equation-of-state parameter is smaller than the
corresponding one of simple holographic dark energy models
\cite{obs3a}, lying inside the phantom regime
($w_0=-1.042^{+0.131}_{-0.124}$). Finally, the left and right
bounds are larger as expected, since the scenario at hand
possesses the additional freedom of varying the gravitational
constant.

For the non-flat geometry, we showed that the scenario at hand is
compatible with observations. We constructed the corresponding
likelihood contours of the  free parameters $\Omega_{\Lambda0}$,
$\Omega_{k0}$, $c$ and $\Delta_G$, taking into account the
uncertainty in $H_0$, and for clarity we presented the
1$\sigma$-bounds  in Table \ref{tab:non_flatcase}. Furthermore,
the best-fit value of the present dark energy equation-of-state
parameter lies inside the phantom regime
($w_0=-1.042^{+0.153}_{-0.204}$), and it is in agreement with the
corresponding one of usual holographic dark energy models
\cite{obs3a}. However, as expected, the additional variation of
the gravitational constant leads to an increase on the extreme
values.

In summary, we conclude that the scenario of holographic dark
energy with  varying gravitational constant can be a candidate for
the description of dark energy.

\begin{acknowledgments}
The data fitting is based on the publicly available {\bf CosmoMC}
package a Markov Chain Monte Carlo (MCMC) code. L. Xu is supported by NSF (10703001), SRFDP (20070141034) of P.R.
China.
\end{acknowledgments}

\appendix*

\section{Observational data and constraints}
\label{Observational data and constraints}

In this appendix we briefly review the main sources of
observational constraints used in this work, namely  Type Ia
Supernovae, Baryon Acoustic Oscillations, Cosmic Microwave
Background $[l_A(z_\ast), R(z_\ast), z_{\ast}]$ and Observational
Hubble Data (OHD).

\subsection{Type Ia Supernovae constraints}

We use the 397 SN Ia Constitution dataset, which includes $397$ SN
Ia \cite{ref:Condata}. Following \cite{ref:smallomega,ref:POLARSKI},
one can obtain the corresponding constraints by fitting the distance
modulus $\mu(z)$ as
\begin{equation}
\mu_{th}(z)=5\log_{10}[D_{L}(z)]+\frac{15}{4}\log_{10}\frac{G_{eff}}{G}+\mu_{0},
\end{equation}
where  $G$ is the current value of effective Newton's constant
$G_{eff}$. In this expression $D_{L}(z)$ is the Hubble-free luminosity distance
$H_0 d_L(z)/c$, with $H_0$ the Hubble constant, defined through the
re-normalized quantity $h$ as $H_0 =100 h~{\rm km ~s}^{-1} {\rm
Mpc}^{-1}$,
 and
\begin{eqnarray}
d_L(z)&=&\frac{c(1+z)}{\sqrt{|\Omega_k|}}sinn[\sqrt{|\Omega_k|}\int_0^z\frac{dz'}{H(z')}], \nonumber\\
\mu_0&\equiv&42.38-5\log_{10}h.\nonumber
\end{eqnarray}
where $sinnn(\sqrt{|\Omega_k|}x)$ respectively denotes
$\sin(\sqrt{|\Omega_k|}x)$, $\sqrt{|\Omega_k|}x$,
$\sinh(\sqrt{|\Omega_k|}x)$ for $\Omega_k<0$, $\Omega_k=0$ and
$\Omega_k>0$.
 Additionally, the observed distance moduli $\mu_{obs}(z_i)$ of SN
Ia at $z_i$ is
\begin{equation}
\mu_{obs}(z_i) = m_{obs}(z_i)-M,
\end{equation}
where $M$ is their absolute magnitudes.

For the SN Ia dataset, the best-fit values of the parameters $p_s$
  can be determined by a likelihood analysis, based on
the calculation of
\begin{eqnarray}
\chi^2(p_s,M^{\prime})\equiv \sum_{SN}\frac{\left\{
\mu_{obs}(z_i)-\mu_{th}(p_s,z_i)\right\}^2} {\sigma_i^2}
=\sum_{SN}\frac{\left\{ 5 \log_{10}[D_L(p_s,z_i)] - m_{obs}(z_i) +
M^{\prime} \right\}^2} {\sigma_i^2}, \ \ \ \ \label{eq:chi2}
\end{eqnarray}
where $M^{\prime}\equiv\mu_0+M$ is a nuisance parameter which
includes the absolute magnitude and the parameter $h$. The nuisance
  parameter $M^{\prime}$ can be marginalized over
analytically \cite{ref:SNchi2} as
\begin{equation}
\bar{\chi}^2(p_s) = -2 \ln \int_{-\infty}^{+\infty}\exp \left[
-\frac{1}{2} \chi^2(p_s,M^{\prime}) \right] dM^{\prime},\nonumber
\label{eq:chi2marg}
\end{equation}
resulting to
\begin{equation}
\bar{\chi}^2 =  A - \frac{B^2}{C} + \ln \left( \frac{C}{2\pi}\right)
, \label{eq:chi2mar}
\end{equation}
with
\begin{eqnarray}
&&A=\sum_{SN} \frac {\left\{5\log_{10}
[D_L(p_s,z_i)]-m_{obs}(z_i)\right\}^2}{\sigma_i^2},\nonumber\\
&& B=\sum_{SN} \frac {5
\log_{10}[D_L(p_s,z_i)]-m_{obs}(z_i)}{\sigma_i^2},\nonumber
\\
&& C=\sum_{SN} \frac {1}{\sigma_i^2}\nonumber.
\end{eqnarray}
Relation (\ref{eq:chi2}) has a minimum at the nuisance parameter
value $M^{\prime}=B/C$, which contains information of the values of
$h$ and $M$. Therefore, one can extract the values of $h$ and $M$
provided the knowledge of one of them. Finally, note that the
expression
\begin{equation}
\chi^2_{SN}(p_s)=A-(B^2/C),\label{eq:chi2SN}\nonumber
\end{equation}
which coincides to (\ref{eq:chi2mar}) up to a constant, is often
used in the likelihood analysis
\cite{ref:smallomega,ref:JCAPXU,ref:SNchi2}, and thus in this case
the results will not be affected by a flat $M^{\prime}$
distribution.

\subsection{Baryon Acoustic Oscillation constraints}

 The Baryon Acoustic Oscillations are detected in the clustering of the
combined 2dFGRS and SDSS main galaxy samples, and measure the
distance-redshift relation at $z = 0.2$. Additionally, Baryon
Acoustic Oscillations in the clustering of the SDSS luminous red
galaxies measure the distance-redshift relation at $z = 0.35$. The
observed scale of the BAO calculated from these samples, as well as
from the combined sample, are jointly analyzed using estimates of
the correlated errors to constrain the form of the distance measure
$D_V(z)$ \cite{ref:Eisenstein2005,ref:Percival2,ref:Percival3}
\begin{equation}
D_V(z)=\left[(1+z)^2 D^2_A(z) \frac{cz}{H(z)}\right]^{1/3}.
\label{eq:DV}
\end{equation}
In this expression  $D_A(z)$ is the proper (not comoving) angular
diameter distance, which has the following relation with $d_{L}(z)$
\begin{equation}
D_A(z)=\frac{d_{L}(z)}{(1+z)^2}.\nonumber
\end{equation}
The peak positions of the BAO depend on the ratio of $D_V(z)$ to the
sound horizon size at the drag epoch (where baryons were released
from photons) $z_d$, which can be obtained by using a fitting
formula \cite{ref:Eisenstein}:
\begin{eqnarray}
&&z_d=\frac{1291(\Omega_mh^2)^{-0.419}}{1+0.659(\Omega_mh^2)^{0.828}}[1+b_1(\Omega_bh^2)^{b_2}],
\end{eqnarray}
with
\begin{eqnarray}
&&b_1=0.313(\Omega_mh^2)^{-0.419}[1+0.607(\Omega_mh^2)^{0.674}] \\
&&b_2=0.238(\Omega_mh^2)^{0.223}.
\end{eqnarray}
In this work we use the data of $r_s(z_d)/D_V(z)$ extracted from
the Sloan Digital Sky Survey (SDSS) and the Two Degree Field
Galaxy Redshift Survey (2dFGRS) \cite{ref:Percival3}, which are
listed in Table \ref{baodata}, with $r_s(z)$ the comoving sound
horizon size
\begin{eqnarray}
r_s(z){=}c\int_0^t\frac{c_sdt}{a}=c\int_0^a\frac{c_sda}{a^2H}=c\int_z^\infty
dz\frac{c_s}{H(z)}
{=}\frac{c}{\sqrt{3}}\int_0^{1/(1+z)}\frac{da}{a^2H(a)\sqrt{1+(3\Omega_b/(4\Omega_\gamma)a)}},
\end{eqnarray}
where $c_s$ is the sound speed of the photon$-$baryon fluid
\cite{ref:Hu1, ref:Hu2, ref:Caldwell}:
\begin{eqnarray}
&&c_s^{-2}=3+\frac{4}{3}\times\frac{\rho_b(z)}{\rho_\gamma(z)}=3+\frac{4}{3}\times(\frac{\Omega_b}{\Omega_\gamma})a,
\end{eqnarray}
and  $\Omega_\gamma=2.469\times10^{-5}h^{-2}$ for $T_{CMB}=2.75K$.
\begin{table}[htbp]
\begin{center}
\begin{tabular}{c|l}
\hline\hline
 $z$ &\ $r_s(z_d)/D_V(z)$  \\ \hline
 $0.2$ &\ $0.1905\pm0.0061$  \\ \hline
 $0.35$  &\ $0.1097\pm0.0036$  \\
\hline
\end{tabular}
\end{center}
\caption{\label{baodata} The observational $r_s(z_d)/D_V(z)$
data~\cite{ref:Percival2}.}
\end{table}

Using the data of BAO in Table \ref{baodata}, the inverse
covariance matrix $V^{-1}$ reads \cite{ref:Percival2}:
\begin{eqnarray}
&&V^{-1}= \left(
\begin{array}{cc}
 30124.1 & -17226.9 \\
 -17226.9 & 86976.6
\end{array}
\right).
\end{eqnarray}
Thus, finally, the $\chi^2_{BAO}(p_s)$ is given as
\begin{equation}
\chi^2_{BAO}(p_s)=X^tV^{-1}X,\label{eq:chi2BAO}
\end{equation}
where $X$ is a column vector formed from the values of theory minus
the corresponding observational data, with
\begin{eqnarray}
&&X= \left(
\begin{array}{c}
 \frac{r_s(z_d)}{D_V(0.2)}-0.190533 \\
 \frac{r_s(z_d)}{D_V(0.35)}-0.109715
\end{array}
\right),
\end{eqnarray}
($X^t$ denotes the transpose).

\subsection{Cosmic Microwave Background constraints}

The CMB shift parameter $R$ is provided by \cite{ref:Bond1997}
\begin{equation}
R(z_{\ast})=\frac{\sqrt{\Omega_m
H^2_0}}{\sqrt{|\Omega_k|}}\mathrm{sinn}\left[\sqrt{|\Omega_k|}\int_0^{z{_\ast}}\frac{dz'}{H(z')}\right],
\end{equation}
where  the redshift $z_{\ast}$ (the decoupling epoch of photons)
is obtained using the fitting function \cite{Hu:1995uz}
\begin{equation}
z_{\ast}=1048\left[1+0.00124(\Omega_bh^2)^{-0.738}\right]\left[1+g_1(\Omega_m
h^2)^{g_2}\right],\nonumber
\end{equation}
and where the functions $g_1$ and $g_2$ read
\begin{eqnarray}
g_1&=&0.0783(\Omega_bh^2)^{-0.238}\left[1+ 39.5(\Omega_bh^2)^{0.763}\right]^{-1}\nonumber \\
g_2&=&0.560\left[1+ 21.1(\Omega_bh^2)^{1.81}\right]^{-1}.\nonumber
\end{eqnarray}
Additionally, the acoustic scale is related to the first distance
ratio and is expressed as
\begin{eqnarray}
&&l_A=\frac{\pi}{r_s(z_{\ast})}\frac{c}{\sqrt{|\Omega_k|}}\mathrm{sinn}\left[\sqrt{|\Omega_k|}\int_0^{z_\ast}\frac{dz'}{H(z')}\right].
\end{eqnarray}

Using the data of $l_A, R, z_{\ast}$ from \cite{ref:Komatsu2008}
which are listed in Table \ref{cmbdata}, and their covariance
matrix of $[l_A(z_\ast), R(z_\ast), z_{\ast}]$ presented in
\cite{ref:Komatsu2008}:
\begin{eqnarray}
&&C^{-1}= \left(
\begin{array}{ccc}
 1.800 & 27.968 & -1.103\\
 27.968 & 5667.577 & -92.263\\
 -1.103 & -92.263 & 2.923
\end{array}
\right),
\end{eqnarray}
we can finally calculate the likelihood $L$ as $\chi^2_{CMB}=-2\ln
L$:
\begin{eqnarray}
&&\chi^2_{CMB}(p_s)=\bigtriangleup d_i[C^{-1}(d_i,d_j)][\bigtriangleup
d_i]^t,\label{eq:chi2CMB}
\end{eqnarray}
where $\bigtriangleup d_i=d_i-d_i^{data}$ is a row vector, and
$d_i=(l_A, R, z_{\ast})$.
 \begin{table}[htbp]
 \begin{center}
 \begin{tabular}{c c   cc   } \hline\hline
 ~ &              5-year maximum likelihood ~~~ & error, $\sigma$ &\\ \hline
 $ l_{A}(z_{\ast})$         & 302.10      & 0.86  & \\
 $ R(z_{\ast})$             &  1.710      & 0.019 & \\
 $ z_{\ast}$                & 1090.04     & 0.93&    \\
 \hline\hline
 \end{tabular}
 \caption{The values of  $ l_{A}(z_{\ast})$, $R(z_{\ast})$, and $z_{\ast}$, from 5-year WMAP results \cite{ref:Komatsu2008}.}\label{cmbdata}
 \end{center}
 \end{table}

%%%%%%%%%%%%%%%%%%%%%%%%%%%%%%%%%%%%%%%%%% OHD
\subsection{Observational Hubble Data constraints}

The observational Hubble data are based on differential ages of
the galaxies \cite{ref:JL2002}. In \cite{ref:JVS2003}, Jimenez
{\it et al.} obtained an independent estimate for the Hubble
parameter using the method developed in \cite{ref:JL2002}, and
used it to constrain the equation of state of dark energy. The
Hubble parameter, depending on the differential ages as a function
of the redshift $z$, can be written as
\begin{equation}
H(z)=-\frac{1}{1+z}\frac{dz}{dt}.
\end{equation}
Therefore, once $dz/dt$ is known, $H(z)$ is directly obtained
\cite{ref:SVJ2005}. By using the differential ages of
passively-evolving galaxies from the Gemini Deep Deep Survey
(GDDS) \cite{ref:GDDS} and archival data \cite{ref:archive1},
Simon {\it et al.} obtained $H(z)$ in the range of $0\lesssim z
\lesssim 1.8$ \cite{ref:SVJ2005}. The twelve observational Hubble
data from \cite{ref:0905} are listed in Table \ref{Hubbledata}.
\begin{widetext}
\begin{table}[htbp]
\begin{center}
\begin{tabular}{c|llllllllllll}
\hline\hline
 $z$ &\ 0 & 0.1 & 0.17 & 0.27 & 0.4 & 0.48 & 0.88 & 0.9 & 1.30 & 1.43 & 1.53 & 1.75  \\ \hline
 $H(z)\ ({\rm km~s^{-1}\,Mpc^{-1})}$ &\ 74.2 & 69 & 83 & 77 & 95 & 97 & 90 & 117 & 168 & 177 & 140 & 202  \\ \hline
 $1 \sigma$ uncertainty &\ $\pm 3.6$ & $\pm 12$ & $\pm 8$ & $\pm 14$ & $\pm 17$ & $\pm 60$ & $\pm 40$
 & $\pm 23$ & $\pm 17$ & $\pm 18$ & $\pm 14$ & $\pm 40$ \\
\hline
\end{tabular}
\end{center}
\caption{\label{Hubbledata} The observational $H(z)$
data~\cite{ref:0905}.}
\end{table}
\end{widetext}
Furthermore, in \cite{ref:0807} the authors used the BAO scale as
a standard ruler in the radial direction, and they obtained three
more additional data: $H(z=0.24)=79.69\pm2.32,
H(z=0.34)=83.8\pm2.96,$ and $H(z=0.43)=86.45\pm3.27$.

 The best-fit values of the model parameters from
observational Hubble data \cite{ref:SVJ2005} are determined by
minimizing
\begin{equation}
\chi_{Hub}^2(p_s)=\sum_{i=1}^{15} \frac{[H_{th}(p_s;z_i)-H_{
obs}(z_i)]^2}{\sigma^2(z_i)},\label{eq:chi2H}
\end{equation}
where $p_s$ denotes the parameters contained in the model,
$H_{th}$ is the predicted value for the Hubble parameter,
$H_{obs}$ is the observed value, $\sigma(z_i)$ is the standard
deviation measurement uncertainty, and the summation runs over the
$15$ observational Hubble data points at redshifts $z_i$.

\end{document}